\documentclass[
    ,final            
  ,numberedheadings 
  ]
  {aipproc}

\layoutstyle{8x11single}

\newcommand{\nubar}{\mbox{$\overline{\nu}$}}
\newcommand{\nue}{\mbox{$\nu_e$}}
\newcommand{\nuebar}{\mbox{$\overline{\nu}_e$}}
\newcommand{\numu}{\mbox{$\nu_{\mu}$}}
\newcommand{\numubar}{\mbox{$\overline\nu_{\mu}$}}

\newcommand{\dcp}{\mbox{$\delta_{CP}$}}

\begin{document}

\title{Summary of Long-Baseline Systematics Session at CETUP*2014}

\classification{14.60.Pq, 13.15.+g}
\keywords      {neutrino oscillation, systematic uncertainty}

\author{Daniel Cherdack}{
  address={Colorado State University}
}

\author{Elizabeth Worcester}{
  address={Brookhaven National Laboratory}
}

\begin{abstract}
A session studying systematics in long-baseline neutrino oscillation physics
was held July 14-18, 2014 as part of CETUP* 2014. Systematic effects from
flux normalization and modeling, modeling of cross sections and nuclear
interactions, and far detector effects were addressed. Experts presented
the capabilities of existing and planned tools. A program of study
to determine estimates of and requirements for the size of these effects
was designed. This document summarizes the results of the CETUP* systematics
workshop and the current status of systematic uncertainty studies in
long-baseline neutrino oscillation measurements.
\end{abstract}

\maketitle

\section{Introduction}
For discovery of CP violation, precise understanding of systematic
uncertainty will be required. Until recently, the expected
level of systematic uncertainty for a long-baseline experiment at Fermilab's
Long-Baseline Neutrino Facility (LBNF)
has been estimated based on experience with existing experiments and estimates of
the performance of the next-generation detectors being designed for experiments at LBNF.
These systematic uncertainties 
are approximated in sensitivity calculations by normalization uncertainties that
cover the uncertainties in all of the models used in the
MC generation chain. These include the flux model,
neutrino interaction cross section models, nuclear models,
and models of the near and far detector response.

At the CETUP* workshop, our goal was to summarize existing knowledge of the the leading systematic effects
and to develop a program of study to determine a quantitative prediction for the size of each effect in
a long-baseline neutrino experiment at LBNF. The ultimate goal of this program is to
perform detailed studies of each source of systematic uncertainty that will
determine detector design and analysis performance requirements to
constrain systematic uncertainty to the projected levels. In this document, the tools that have
been developed to study systematic uncertainty in LBNE are described in Section~\ref{sect:tools},
the status of existing studies is described in Section~\ref{sect:current}, and plans for future studies are
described in Section~\ref{sect:planned}. The studies presented here were performed in the context of LBNE, but are
largely applicable to the new international experimental collaboration
at LBNF~\cite{elbnf_loi}.

\section{Workshop Details}
The CETUP* 2014 systematics workshop had 21 registered participants, some of whom
were participating in the concurrent Near Detector workshop and some of whom
were focused exclusively on the systematics workshop. Additionally, several experts
who were unable to attend in person gave presentations over the phone.
Each day of the workshop focused on a particular
source of systematic uncertainty. Table \ref{table:agenda} is the agenda
for each day. Each day, there were approximately two hours of
presentations with most of the day reserved for working time in
which participants discussed issues arising from the presentations and
developed a list of specific studies that are required to understand each
systematic effect. Requirements for development of tools and acquisition
of external input were also determined. Detector requirements were determined
when possible, though generally these will be an output of studies that
have not yet been performed.
\begin{table}
\centering
\caption{Agenda of CETUP* Systematics Workshop}
\begin{tabular}[c]{clll} 
Day  & Topic & Presentation & Speaker \\ \hline
1 & Introduction   & Welcome & Barbara Szczerbinska  \\
  &                & Workshop Goals and Charge & Elizabeth Worcester \\
  &                & Tools for Studying Systematic Uncertainty & Dan Cherdack \\
  & Flux           & Flux Systematic Uncertainties & Paul Lebrun \\
  &                & ND Flux Constraints & Xinchun Tian \\
  &                & Constraining the NuMI Flux & Debbie Harris \\ \hline
2 & Cross sections & ND Cross-Section Constraints (Fast MC) & Xinchun Tian \\
  &                & ND Flux and Cross-Section Constraints (VALOR) & Costas Andreopoulos \\
  &                & Hadronization (GENIE/VALOR) & Costas Andreopoulos \\
  &                & FD Cross-Section Constraints (Fast MC) & Dan Cherdack \\ 
  &                & Systematic Effects from QE, MEC, and RES & Rik Gran \\ \hline
3 & Nuclear Model  & Near Detector Constraints & Roberto Petti \\
  &                & GENIE vs GiBUU & Mindy Jen \\
  &                & Reducing Energy Resolution Uncertainty & Ulrich Mosel \\
  &                & GENIE FSI Model/Systematics & Steve Dytman \\ \hline
4 & Far Detector   & LAr Simulations and Reconstruction & Tom Junk \\
  &                & Constraints from Test Beam Experiments & Xin Qian \\
  &                & Alternative FD Designs: Implications for Systematics & Nuno Barros \\
  &                & Tour of SURF 4850-L & \\ \hline
5 & Discussion     &                     & \\ \hline
\end{tabular}
\label{table:agenda}
\end{table}

\section{Tools}
\label{sect:tools}
The work presented here makes use of a number of tools, some of which are commonly used in
the HEP community and others of which have been developed by the LBNE collaboration specifically
for analysis of long-baseline systematics and sensitivity calculations.
GLoBES\cite{globes1,globes2} is a software package developed to calculate
sensitivities for neutrino oscillation experiments.
The LBNE Fast Monte Carlo (Fast MC) is a simulation suite, combining flux simulations from
G4LBNE (a GEANT4\cite{geant4}-based beam simulation), the GENIE\cite{genie} 
neutrino event generator, and a parameterized detector response
used to simulate the energy deposition of each final-state particle. The
ND Fast MC is an implementation of a fine-grained tracker (FGT) near detector (ND) in the
Fast MC framework. My GLoBES Tools (MGT) is an analysis
framework using GLoBES libraries that calculates experimental sensitivity, including
the effect of systematic variations, using inputs from a simulation such as the Fast MC. 
VALOR is a fitting framework that performs a maximum likelihood fit of near
detector data to constrain systematic uncertainties and then applies these
constraints to oscillation parameter fits of far detector event samples. LArSoft\cite{Church:2013hea}
is a full detector response simulation and reconstruction package for liquid
argon TPCs, used by multiple experiments. Reconstruction algorithms
within LArSoft are under development. Once further development is complete,
the LArSoft simulation will be used in conjunction with with the same flux and cross section
simulations as the current Fast MC to provide full simulations of a LArTPC far
detector at LBNF.

\section{Current Results}
\label{sect:current}

\subsection{GLoBES Studies}
Expected systematic uncertainties on the $\nu_e$ appearance samples in the three-flavor fit for LBNE
are extrapolated from the current performance of the MINOS~\cite{Adamson:2013ue,Adamson:2011qu}
and T2K~\cite{t2k_2013} experiments, as shown in Table~\ref{tab:nuesysts}.
\begin{table}[!hb]
  \caption {The dominant systematic uncertainties on the $\nu_e$ appearance 
    signal prediction in MINOS and T2K and a conservative projection of the 
    expected uncertainties in an experiment at LBNF. In each case, the quoted uncertainty is
    the effect on the $\nu_e$ appearance signal only. These uncertainties 
    are the \emph{total} expected uncertainties on the $\nu_e$ appearance signal 
    which include both correlated and uncorrelated uncertainties in the 
    three-flavor fit.\vspace{2pt}}
\label{tab:nuesysts}
\begin{tabular}{l|c|c|c|l} \hline
Source of & MINOS & T2K & ELBNF & Comments \\
Uncertainty & $\nu_e$ & $\nu_e$ & $\nu_e$ & \\ \hline
\multicolumn{5}{c}{Flux Determination}  \\ \hline
Beam Flux & 0.3\% & 2.9\% & 2\% & MINOS is normalization only.\\ 
after N/F & & & & ELBNF normalization  and shape   \\
extrapolation & & & & highly correlated between $\nu_\mu/\nu_e$. \\ \hline
\multicolumn{5}{c}{Neutrino interaction modeling}  \\ \hline
Simulation & 2.7\% & 7.5\% & $\sim 2\%$ & Hadronization models are better  \\
includes: & & & & constrained in the ELBNF LArTPC. \\
hadronization & & & &  N/F cancellation larger in MINOS/ELBNF. \\ 
cross sections & & & & X-section uncertainties larger at T2K energies. \\ 
nuclear models & & & & Spectral analysis in ELBNF provides \\ 
& & & & extra constraint. \\ \hline
\multicolumn{5}{c}{Detector effects}  \\ \hline
Energy scale  & 3.5\% & included& (2\%) & Included in ELBNF $\nu_\mu$ sample  \\ 
($\nu_\mu$) & & above & &  uncertainty only in 3-flavor fit. \\
& & & & MINOS dominated by hadronic scale. \\ \hline
Energy  scale & 2.7\% & 3.4\% & 2\% & Totally active LArTPC with calibration \\
($\nu_e$) & & includes & & and test beam data lowers uncertainty. \\
 & & all FD & & \\
 & & effects & & \\ \hline 
Fiducial & 2.4\% & 1\% & 1\% & Larger detectors = smaller uncertainty. \\ 
volume & & & & \\ \hline\hline
Total  & 5.7\% & 8.8\% & 3.6 \% & Uncorrelated $\nu_e$ uncertainty in  \\ 
& & & & full ELBNF 3-flavor fit = 1-2\%. \\ \hline
\end{tabular}
\end{table}
The sensitivity of long-baseline oscillation parameter measurements 
has been evaluated in GLoBES fits that account for statistical uncertainties, 
oscillation parameter uncertainties, and these non-oscillation systematic 
uncertainties. The latter is accomplished via eight normalization parameters; 
one each for the signal and background of four simultaneously fit analysis 
samples ($\nue$, $\nuebar$, $\numu$, $\numubar$).
These eight parameters are assumed to be completely uncorrelated, 
as any correlated uncertainty is expected to cancel. Estimates for 
these normalization uncertainties are 5\% (1\%) signal  and 10\% (5\%) 
background for the $\numu$ ($\nue$) samples. These values 
reflect the uncorrelated portion of the total uncertainty projections for
each sample, including constraints from external and near detector data.
The portion of the systematics uncertainties that are correlated 
amongst analysis sample are expected to largely cancel in combined fits.

\begin{figure}[!htb]
\centering
\includegraphics[width=0.45\linewidth]{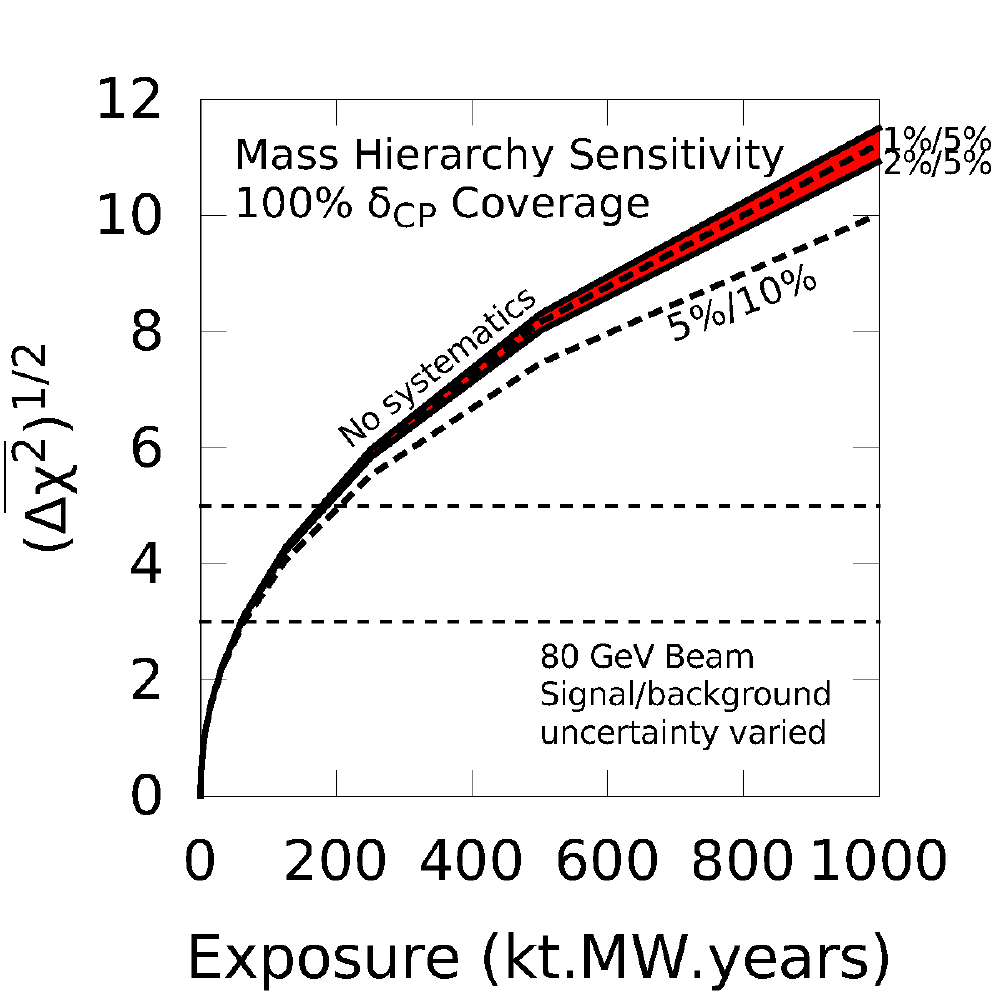}
\includegraphics[width=0.45\linewidth]{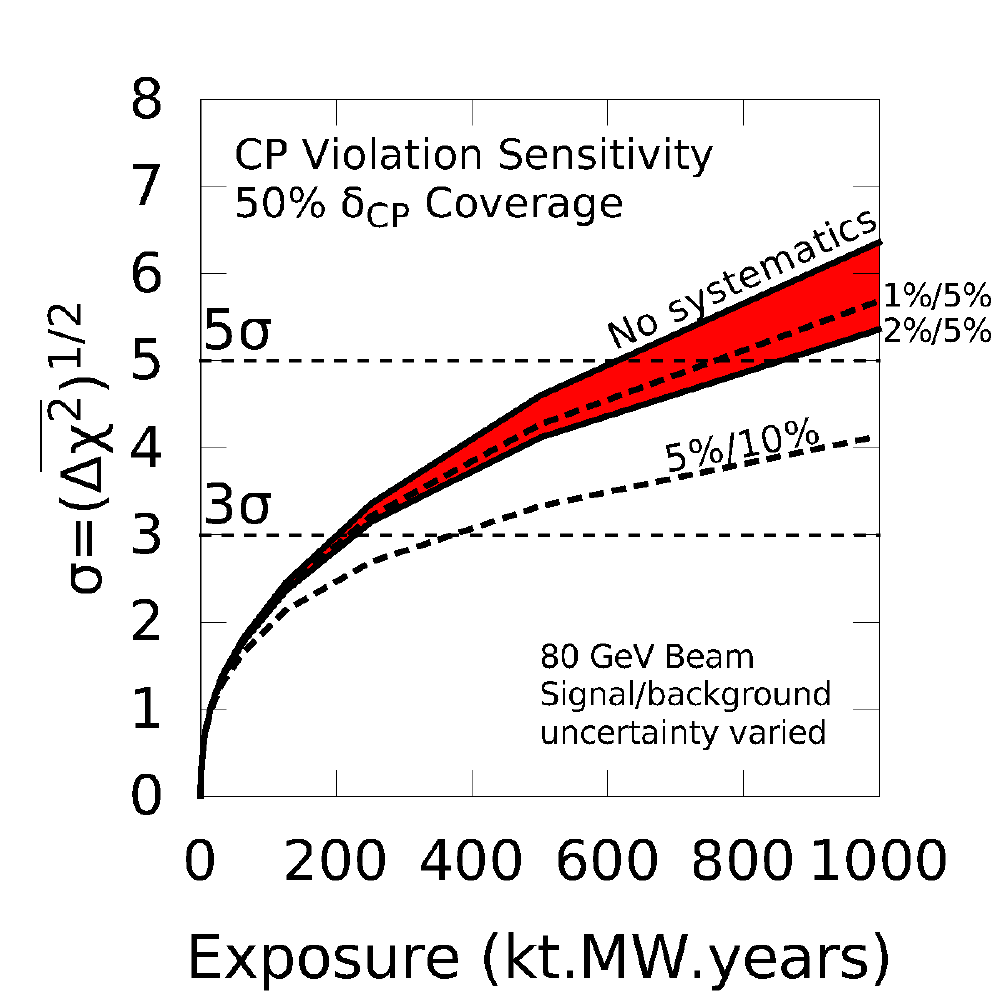}
\caption{Expected sensitivity of an experiment at LBNF to determination of the neutrino mass
  hierarchy (left) and discovery of CP violation, i.e. $\dcp \ne$ 0 or $\pi$,
  (right) as a function of exposure in kt-MW-years, assuming 
  equal running in neutrino and antineutrino mode, for a range of values for
  the residual $\nue$ and $\nuebar$ signal and
  background normalization uncertainties. The sensitivities quoted
  are the minimum sensitivity for 100\% of $\dcp$ values in the case of 
  mass hierarchy and 50\% of $\dcp$ values in the case of CP violation.
  Sensitivities are for true normal hierarchy; neutrino mass hierarchy is assumed to
  be unknown in the CPV fits. Values of the oscillation parameters and their
  uncertainties used in this calculation are taken from \cite{Gonzalez-Garcia:2014bfa}.}
\label{fig:exp_systs}
\end{figure}

The sensitivity of the mass hierarchy determination and the CP
violation measurement to possible values for
the uncorrelated $\nu_e$ signal and background normalization uncertainties
in the GLoBES-based calculation are shown in Fig.~\ref{fig:exp_systs}.
The impact of systematic uncertainty on the CP violation sensitivity
is clear; the normalization of the $\nue$ sample,
relative to the $\nuebar$, $\numu$, and $\numubar$ samples after all constraints from
external, near detector, and far detector data have been applied, must be determined 
at the 1-2\% level in order to reach 5$\sigma$ sensitivity for exposures less 
than 900 kt-MW-years. An important goal of the long-baseline community and the
CETUP*2014 systematics workshop is to individually evaluate the effect of individual systematic
uncertainties from flux determination, neutrino interaction cross section models, nuclear models,
and near and far detector response rather than relying on these expected normalization uncertainties.

\subsection{Flux}
The uncertainties from the generation
of hadrons that decay into neutrinos, which are produced off the target 
and by secondary and tertiary interactions in the target hall and the decay pipe, 
are known to be large ($\sim$10\%) and are the leading source of uncertainty in 
conventional neutrino beamlines. Studies of ND constraints are the best handle
on how the resulting flux uncertainties will be propagated to FD measurements.

To evaluate the capabilities of proposed near detectors to
constrain the flux, initial studies have been performed for a 
FGT ND design making use of the ND Fast MC.
These studies demonstrate the capability of a FGT to constrain the
absolute flux using fully-leptonic interaction channels,
as well as the flux shape and the FD/ND flux ratio.
Initial studies predict a 2\% statistical 
uncertainty for neutrino-electron scattering which can be used to 
constrain the flux below $\sim$5~GeV, and a 3\% statistical uncertainty 
for inverse muon decay, which can be used to measure the flux above $\sim$10~GeV.
The low-$\nu_0$ method~\cite{srmishra-reviewtalk,Adamson:2009ju,Bodek:2012cm}
can be used to constrain the shape of the flux
as well as FD/ND shape differences and is expected to do so at the 1-2\% level.

\begin{figure}[!htb]
  \includegraphics[width=\linewidth]{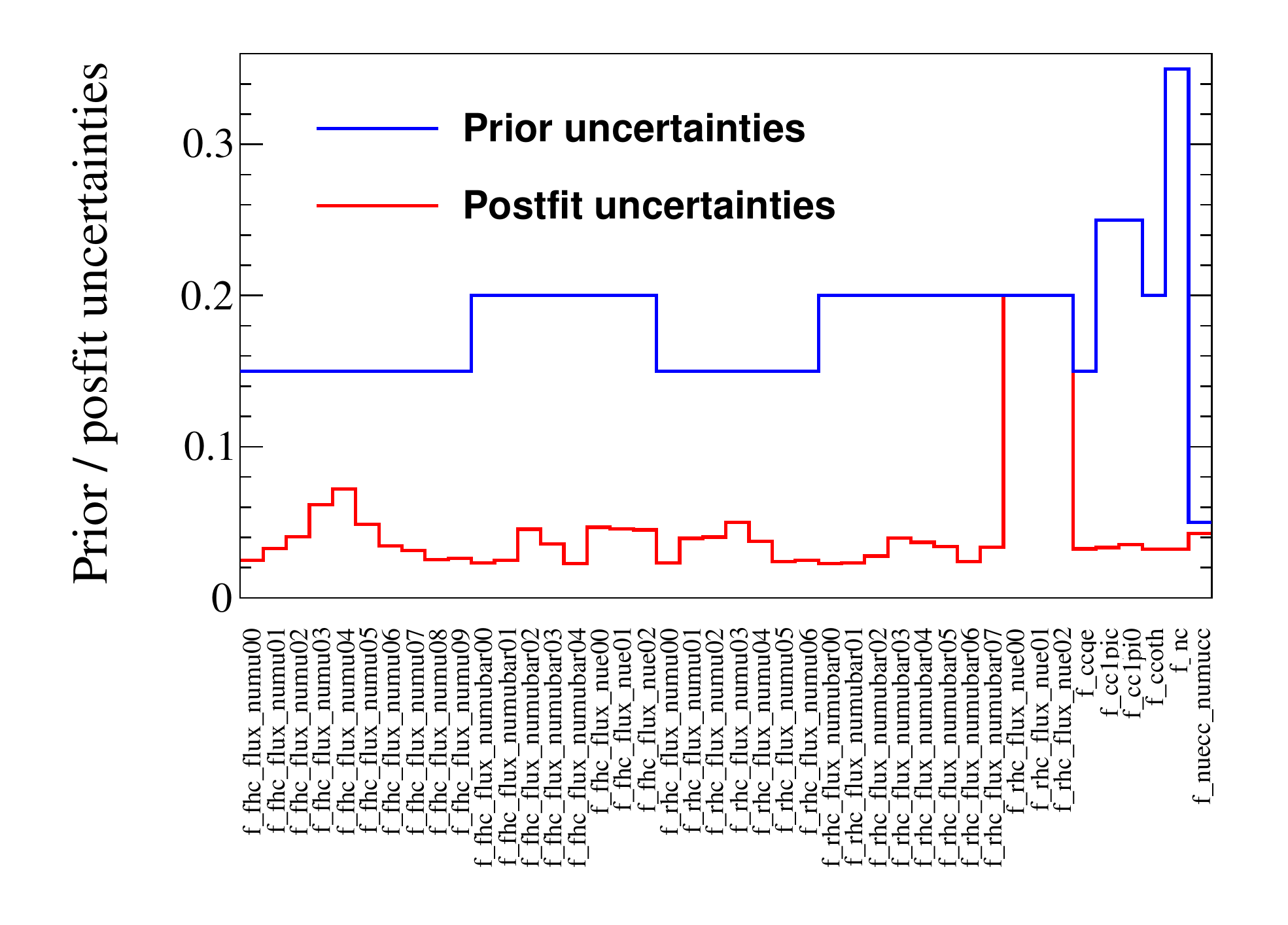}
  \caption {
  Example of prior uncertainties (blue) compared with post-fit uncertainties (red) on binned flux
  normalization parameters from a VALOR 
  fit of simulated near-detector data. In this fit, the true values of all the FHC $\nu_{\mu}$ flux parameters are
  set to be 10\% greater than nominal and 
  the true values of the other parameters are set to be nominal. No statistical fluctuations are applied.
  Near-detector efficiency parameters are included in the near-detector fit but are then marginalized.
  The post-fit uncertainties in the RHC $\nu_{e}$ flux parameters are identical to the prior uncertainties
  since there is no RHC $\nu_{e}$ sample included in this fit, which means that these parameters are not
  constrained by this fit.
  }
  \label{fig:covmat_diagonal_with_prior}
\end{figure}

Studies using the VALOR analysis technique to constrain simulated data derived from the Fast MC
event samples also demonstrate the significant ability of a FGT ND to constrain the flux rate and shape.
As shown in Fig.~\ref{fig:covmat_diagonal_with_prior}, 
the post-fit uncertainty 
in most flux bins for a sample VALOR fit is less
than 5\%, which is consistent with the uncorrelated $\nu_\mu$ signal normalization
uncertainty assumed by the sensitivity calculations. 
Since the flux for the $\nue$($\nuebar$) appearance
and the $\numu$($\numubar$)-disapearance analysis samples are identical,
the unceratinties are 100\% correlated;
i.e., the $\numu$($\numubar$) sample constrains the flux for the $\nue$($\nuebar$) sample.

Systematic uncertainty in the far detector flux arising
from uncertainties in positions of beamline elements induce significant
reduction in sensitivity in the absence of near detector
constraints, but preliminary studies indicate that the near detector will
significantly constrain the flux, so these effects are not expected
to be a leading source of systematic uncertainty. Furthermore, these
uncertainties have been evaluated using beam simulations and
the quadrature sum of the uncertainty propagated to the ND/FD ratio
is determined to be at the 1\% level. 
\subsection{Cross Sections and Nuclear Interactions}

\begin{figure}[!htb]
\centering
\includegraphics[width=0.7\linewidth]{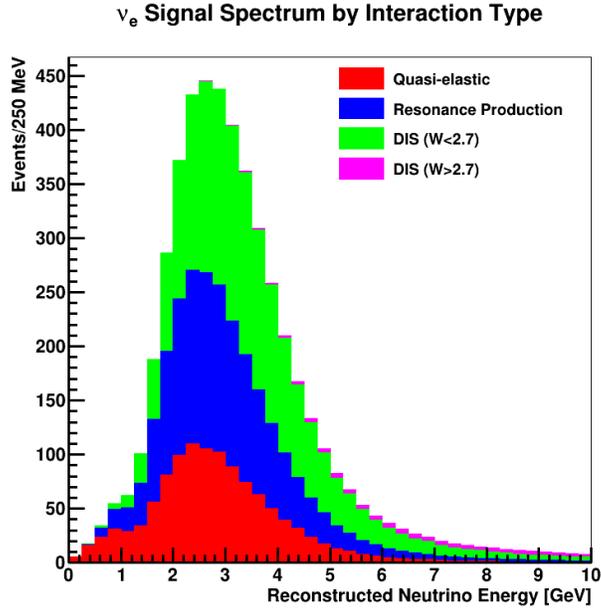}
\caption{Example $\nu_e$ appearance spectrum at LBNF. Events are separated by interaction
type using Fast MC signal selection criteria.}
\label{fig:inttype}
\end{figure}

\begin{figure}[!htb]
\centering
\includegraphics[width=0.9\linewidth]{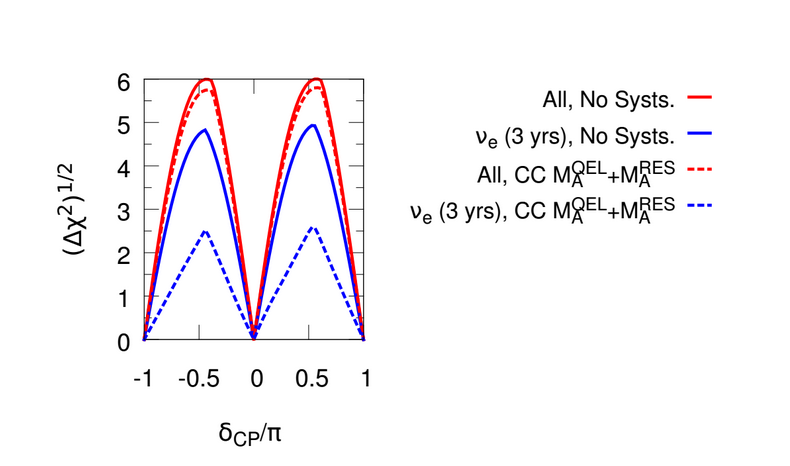}
\caption{The CP violation sensitivity for the nominal 245 kt-MW-years
of exposure calculated using inputs from the Fast MC in a fit to all
four ($\nu_e$, $\overline\nu_e$, $\nu_{\mu}$, $\overline\nu_{\mu}$)
samples (red) and a fit to the $\nu_e$ appearance sample only (blue),
for the case of no systematic uncertainty (solid) and the case in
which both $M_A^{QE,CC}$ and $M_A^{RES,CC}$ are allowed to vary with a
1$\sigma$ uncertainty of 20\% (dashed).}
\label{fig:MAresqesyst}
\end{figure}

As seen in Fig.~\ref{fig:inttype}, at LBNF the far detector event samples will contain
significant contributions from quasielastic,
resonance production, and deep inelastic scattering (DIS) interactions, so it is
important to understand systematics arising from uncertainty in the 
cross-section models for each of these neutrino interaction processes. Results
from MGT using Fast MC inputs show that a fit to all four far detector samples
($\nu_e$, $\overline\nu_e$, $\nu_{\mu}$, $\overline\nu_{\mu}$) significantly
constrains cross-section systematic uncertainty even in the case where multiple
cross-section parameters are allowed to vary simultaneously within their
GENIE uncertainties. For example,
fits are performed in which $M_A^{QE}$ and $M_A^{RES}$ for charged current (CC)
interactions are both allowed to vary within their GENIE uncertainties of $\sim\pm$20\%.
This essentially allows the constituent sample normalizations in the reconstructed energy
spectra to vary by $\sim$20\%. As seen in the example fit shown in Fig.~\ref{fig:MAresqesyst}, 
this level of allowed variation results in a dramatic reduction in sensitivity if one fits
only the $\nu_e$ appearance signal without constraint from the 
$\nuebar$ and $\numu$/$\numubar$ samples.
In contrast, for a four-sample fit, the same level of allowed fluctuations lead to 
a significantly smaller degradation of the CPV sensitivity. Comparing the fraction
of $\dcp$ values for which a 3$\sigma$ discovery of CP violation can be made with an
exposure of 245 kt-MW-years, i.e., 3$\sigma$ CPV coverage, between the two fits,
we find a 54\% reduction in 3$\sigma$ CPV coverage for the $\nu_e$ appearance-only fit is
reduced to a 2\% reduction in 3$\sigma$ CPV coverage for the four-sample fit.
This result includes a 10\% uncertainty in the $\nu/\nubar$
cross-section ratio and a 2.5\% uncertainty in the $\nue/\numu$
cross-section ratio.
Preliminary studies from VALOR and the ND Fast MC
demonstrate significant constraint on cross-section systematics from the 
near detector as well. 

Nuclear models enter into the simulation of neutrino interactions both through modeling of
initial-state interactions, i.e., interactions between the initial state of the nucleons
and virtual particles within the nucleus with the neutrino, and modeling of 
final-state interactions (FSI), i.e., interactions of the particles exiting the primary 
interaction vertex with the nuclear medium.
Uncertainties in initial-state interactions
due to naive modeling of the environment of the nucleus have thus far been taken
into account through inflation of the uncertainties on the free nucleon or quark interaction
model, such as $M_{A}^{QE}$. Final-state interactions can alter
event reconstruction in two distinct ways. The first is a smearing of the total energy
available to be deposited in the detector. The second is the misidentification of 
event topologies used to classify the neutrino flavor and interaction mode.
Uncertainties in selection efficiencies and event-sample
migrations due to intranuclear rescattering can be studied 
with the Fast MC and have been implemented in VALOR. 

\subsection{Detector Effects}
Methods to estimate far detector properties, such as single-particle energy
resolutions, missing hadronic energy, energy-scale uncertainties,
event selection efficiencies, and background rejection rates have been 
implemented in the Fast MC. The parameters that control these effects 
are determined by GENIE event kinematics, typical event selection criteria, 
ICARUS\cite{Amoruso:2003sw,Ankowski:2006ts} and T2K\cite{T2K2kmprop}
results, and LArSoft-based simulations. Current Fast MC
sensitivity calculations reflect the nominal input values for the 
energy resolutions and missing energy. 

\section{Planned Work}
\label{sect:planned}
At the CETUP*2014 systematics workshop, we developed a list of studies that would inform the study of
systematic uncertainty in a long-baseline neutrino oscillation analysis. In this section, we describe
the status of in-progress and planned studies. In many cases, this work is a collaboration among
multiple experimental efforts, phenomonologists, and simulation groups. An important part of this
program is support for the efforts of phenomonologists and simulation groups to implement a range
of models into neutrino event generators for comparison to experimental results.

\subsection{GLoBES Studies}
The developers of GLoBES have implemented more sophisticated treatment of
systematic uncertainty into an unreleased version of the software, which 
has been used in \cite{Coloma:2012ji,Coloma:2013rqa,Coloma:2013tba}. 
This updated version of GLoBES has been made available to LBNE collaborators
and is being incorporated into future sensitivity studies.

\subsection{Flux}
Initial studies of neutrino-electron scattering, inverse muon decay,
and the low-$\nu_0$ method, which currently predict absolute flux 
normalization at the 2-3\% level and relative normalization and ND/FD
ratio at the 1-2\% level,
must be followed up with more detailed treatment of systematic uncertainties
affecting these results. For example, event selections for purely lepton interactions 
rely heavily on the angle of the final-state lepton, and uncertainties of 
the angular dispersion of the beam must be included to validate the current results.
Such studies are planned and in progress. 

It has been proposed that coherent pion production,
which has the same cross section from neutrinos and antineutrinos and 
does not suffer from nuclear effects, will help to constrain the 
$\nu/\bar{\nu}$ flux ratio.
Alternate models of coherent production\cite{BS1,AR1,AR2,AR3}, which are known to
vary greatly in their prediction of the coherent interaction cross sections,
are expected to be included in an upcoming GENIE release, allowing for detailed study of
this proposal.

More work is needed to asses the 
impact of hadronization model uncertainties, which are the leading source 
of uncertainty in the flux prediction.
Minerva\cite{Osmanov:2011ig} measurements of the NuMI flux have been obtained in the form of
a flux correlation matrix\cite{Higuera:2014azj}; flux uncertainties based on this data
are being implemented in the Fast MC and will be propagated to LBNF analyses.
Since the NuMI and LBNF beams share many key properties,
this technique should provide reasonably accurate results.
A flux driver with Minerva flux re-weighting tools is also
being implemented in GENIE to allow
reweighting of events based on the set of interactions that produced the
neutrino. Once these tools are available,
additional studies will be performed to determine
the precision with which the absolute flux, flux shape, and ND/FD flux
ratio can be measured given realistic uncertainties in the hadronization model.

Extrapolation of the flux at the near detector, measured using the 
methods described above, to the far detector will require understanding of 
the relative energy scale between the near and far detectors. To determine 
the required precision on the relative energy scale, which we assume will be 
dominated by the far detector energy reconstruction, we will explore the 
variation in physics sensitivity for a range of values for muon energy 
bias and resolution, hadron energy resolution, and energy reconstruction 
bias from undetected and mis-identified particles. Undetected particles include neutrons
and particles below threshold. 

\subsection{Cross Sections and Nuclear Interactions}

Current studies based on Fast MC samples address cross-section 
parameter variations for which GENIE reweights are available. 
To evaluate cross-section uncertainties beyond those considered by GENIE,
we will make comparisons between observables from the Fast MC generated 
using the nominal GENIE and with alternate versions of GENIE. These alterations 
include adjusting non-reweightable parameters and the use of additional and 
alternate cross-section models.
In particular, models that describe initial-state interactions,
including models of long- and short-range correlations amongst
nucleons\cite{Benhar1994493,Jen:2014aja},
random phase approximation (RPA) effects,
meson-exchange currents (MEC),
2p-2h effects in CCQE interactions\cite{Nieves:2004bh,Martini:2009uj,Gran:2013kda,Nieves:2013fr}, and
extensions of these models to resonance production interactions must be implemented. The effective spectral
function model\cite{Bodek:2014pka} includes all of these effects by tuning to electron
scattering data.
These models are all in various development stages for GENIE.
Implementations of the Alvarez-Ruso\cite{AR1,AR2,AR3} and Berger-Sehgal\cite{BS1} models of
coherent pion production are also expected in upcoming versions of GENIE.
These models can be compared
with the fluctuations that are allow by uncertainties in effective parameters
like $M_{A}^{QE}$, which have subsumed the uncertainties induced by nuclear effects.
We can also compare systematic error coverage against recent data (e.g. Minerva results)
and alternate generators (e.g. NuWro\cite{nuwro} and GiBUU\cite{Buss:2011mx}).

An alternate model for DIS interactions\cite{Kulagin:2007ju} will also be implemented in GENIE
and predictions from this new model will be compared to the current implementation based on
Bodek-Yang to determine
if the two models are consistent within their uncertainties. Models of DIS interactions
predict only the bulk properties of the final-state hadronic system; other models are
required to predict the particle content of the hadronic system as well as the momenta of 
each final-state hadron. In GENIE this is done with the AGKY model, which is
difficult to reweight. To test the uncertainties in the AGKY model, new GENIE samples
are generated in which all of the tunable 
parameters are adjusted by their $\pm1\sigma$ uncertainties; any parameter variation
resulting in significant changes to the energy spectra will be studied further
using a parameterized reweight function.
Uncertainties in the formation length, the distance a hadron travels within the nucelus before i
t can interact with the nuclear medium, will also be studied.

Studies to determine if the reweightable parameters in the 
GENIE intranuclear rescattering 
model adequately cover the model uncertainties will come from comparisons 
with external data, eg:\cite{2011AIPC.1405..213D,Eberly:2013/05/01kqa,Walton:2014esl},
and comparisons with alternate models, in this case GiBUU. Additionally, tests in which
tunable, but currently non-reweightable, parameters in the GENIE intranuclear rescattering
model are fluctuated within reasonable ranges will be required to 
determine if they will lead to uncertainties on oscillation parameter measurements.
The effect of uncertainty in the amount of undetected energy is being 
studied in the Fast MC and will be used to determine the neutron response 
calibration required to achieve the necessary energy resolution for accurate determination
of CPV. Measurements and simulations by CAPTAIN\cite{Berns:2013usa} will be extrapolated
to determine if the required level of energy resolution is possible
with the current FD design.

There is concern that it will be difficult to isolate nuclear reinteraction
uncertainties from flux, cross-section, and detector model uncertainties,
especially with regard to $\nu/\nubar$ differences. This could lead to differences
in the $\nue$ and $\nuebar$ appearance energy spectra that could be mistaken 
for CPV. A method to combat this by comparing the $\nue$ and $\nuebar$ samples to the $\numu$ and $\numubar$ samples,
where CPV effects in oscillation probabilities are negligible, has been suggested.
However, this requires that the hadronic final states in the appearance and disappearance
samples are comparable. Further work is needed to
develop this method and evaluate the potential reduction in related systematic uncertainties.

\subsection{Far Detector Effects}
Propagation of uncertainty in energy scale and energy resolution
inputs into the Fast MC sensitivity calculations is in progress.
The initial plan of 
study is to determine required constraints on these effects using the Fast MC
and to interact with the various near-term short-baseline neutrino
and test-beam experiments to 
ensure that the required constraints will be available. 
In the longer term, more sophisticated
simulation and analysis algorithms will be included in the Fast MC and the
requirements and sensitivities will be re-evaluated as more information
becomes available.

Development of a MicroBooNE\cite{Chen:2007ae} configuration
of the Fast MC and LArSoft studies of
single particle response will allow for some level of validation of the
single-particle energy
resolutions, missing hadronic energy, energy-scale uncertainties,
event selection efficiencies, and background rejection rates
which are input to the Fast MC. Future improvements
to these estimates will come from data and data-MC comparisons from
the LBNE 35-t prototype, the CERN Neutrino Platform\cite{cern_neutrino_platform}
prototypes, MicroBooNE, LAr1-ND\cite{Adams:2013uaa},
LArIAT\cite{Cavanna:2014iqa}, and CAPTAIN.

\section{Conclusion}
The CETUP* 2014 systematics session was a valuable opportunity for detailed
and productive discussion of systematic uncertainty in long-baseline physics
measurements. 
The primary result of the workshop is an understanding of available tools, the
present status of uncertainty studies, and a detailed plan of further study that is
needed. A summary of these results is provided in this document; detailed reports
on specific topics are available in other contributions to the CETUP* 2014 proceedings.


\begin{theacknowledgments}
We would like to thank
CETUP* (Center for Theoretical Underground Physics and Related Areas) for its hospitality
and partial support during the 2014 summer program. In addition, we would like to thank
the CETUP* organizers, Baha Balantekin and Barbara Szczerbinska, for their efforts to
organize this valuable opportunity for collaboration and all of the
speakers and participants in the CETUP* 2014 systematics session for their contributions.
This work is supported in part by the United States Department of Energy, including
contract DE-SC0012704.
\end{theacknowledgments}



\bibliographystyle{aipproc}   

\bibliography{cetup_proc_bib}

\IfFileExists{\jobname.bbl}{}
 {\typeout{}
  \typeout{******************************************}
  \typeout{** Please run "bibtex \jobname" to optain}
  \typeout{** the bibliography and then re-run LaTeX}
  \typeout{** twice to fix the references!}
  \typeout{******************************************}
  \typeout{}
 }

\end{document}